\documentclass[12pt]{article}

\usepackage{amssymb}
\usepackage{amsmath}
\usepackage{epsfig}
\usepackage{subfigure}
\usepackage{mathrsfs}

\begin{document}

\title{Properties of Black Hole Radiation from Tunnelling}
\author{{Timothy Clifton\thanks{
e-mail: TClifton@astro.ox.ac.uk}}\\
{\small {\textit{Department of Astrophysics,}}}\\
{\small {\textit{University of Oxford, Oxford OX1 3RH, UK}}}}
\date{{\normalsize {\today}}}
\maketitle

\begin{abstract}

We consider the space-time associated with the evaporation of a black
hole by quantum mechanical tunnelling events.  It is shown that the
surface through which tunnelling occurs is distinct from the global
event horizon, and that this has consequences for the radiation
reaching future null infinity.  A spherical collapse process is
modelled, and the radiation expected to be observed at future null
infinity is calculated.  It is shown that external observers witness
an evaporation process that begins as the tunnelling surface is
exposed, and ends as the collapsing object passes behind its event
horizon.  The sensitivity of emitted radiation to the
collapse process is illustrated.

\end{abstract}

\section{Introduction}

A common interpretation of Hawking radiation \cite{hawk1} is that of virtual
particles tunnelling through the horizon.  In this picture a virtual
pair is nucleated just inside (or just outside) the horizon.  A
quantum mechanical event then transports the positive energy
particle from inside the horizon to the outside (or the negative
energy particle from outside to inside).  The result is a
real particle materialising outside the black
hole, and the mass of the hole being reduced.  This
long considered heuristic explanation of Hawking radiation was recently
given a solid basis by Parikh and Wilczek \cite{par1}.  The
calculation they performed took into account the changing geometry of
the space-time as the tunnelling events occur, and provides a natural time
coordinate with which to parameterise the mass loss of the hole, as
well as giving the location of the origin of radiation.  Interestingly,
it was shown that the spectrum of emitted radiation is not a precise
black-body, as had previously been shown in \cite{kraus}, although it was not made clear how this
could be linked to the collapse process that formed the hole.

Here we consider a collapse process leading to the formation and
subsequent evaporation of a
black hole.  We model this as a collapsing spherical shell of matter.
Once formed, the hole is assumed to decrease in mass by the
tunnelling process described above.  The resulting geometry (as
previously considered by Brown and Lindesay \cite{lind1, lind2, lind3})
exhibits a global event horizon that is displaced from the origin of
the radiation.  The energy-momentum of the escaping radiation is then
calculated.  It is found, as expected, that an asymptotic observer watching the
collapse witnesses an increase in the flux of radiation as he/she is
exposed to the surface through which tunnelling occurs.  This
appears to the external observer as if radiation is being emitted
before the collapsing object passes its own event horizon.  The flux
of radiation then continues to increase until the event horizon is reached.  Similar results, by an alternative
method, have been found in \cite{vach1}.

It is of interest to determine the sensitivity of the emitted radiation
to the collapse process that forms the black hole.  As
is well known, all information about the collapse is removed in
the limit that the event horizon is approached.  Only the early
radiation is sensitive to it \cite{dav1}.  In the case of a
static black hole it is unclear exactly what is meant by `early'
(i.e. if this means a short time before the end of evaporation, or in the
remote past).  In the present case we have both a beginning and an end to the
evaporation process: when the tunnelling surface is exposed and when
the black hole evaporates completely.  This allows an
investigation of whether the radiation emitted between these times can
contain any information about the structure that collapsed to form the
hole.  That the tunnelling surface is known to be separate from
the global event horizon indicates that this may be the case, as
the modes that are infinitely suppressed as the horizon is approached can
then be non-zero.

In section 2 we consider the geometry of the system.  The mass of the
hole is made to decrease as a function of the time coordinate used in
the tunnelling calculation \cite{par1} (the proper time of a
freely falling observer).  The
resulting line-element is then transformed into double null coordinates, as
in \cite{lind2}.  In section 3 we discuss
the separation of the global event horizon from the surface $r=2m$.
The tunnelling calculation suggests that the outgoing radiation is
emitted from $r=2m$, and not the global horizon.  This is in good
agreement with the results of \cite{visser,di,niel1}. Section 4 contains
a calculation of the renormalised vacuum expectation value of the
energy-momentum tensor in the two dimensional analogue of this space-time, as prescribed by the point splitting
method of Davies and Fulling \cite{dav2}.  The
corresponding energy-momentum tensor in the Vaidya space-time has been
considered in \cite{his, bal}.   In section 5 we
use the results of the previous calculations to consider the
radiation measured by an external observer watching a spherical
collapse, and show
the extent to which this radiation is effected by the process that formed
the hole.  A discussion is provided in section 6.

\section{Geometry of the system}

A natural choice for the parameterisation of tunnelling events through
the horizon is the proper time of a radially in-falling observer.  This
is given by the Panlev\'{e} time \cite{pan}, or river time, $t$, used by Parikh
and Wilczek in their calculation of the rate of these events \cite{par1}.  For a
black hole geometry with constant mass, this time coordinate is related to the
more usual Schwarzschild time, $t_S$, by
\begin{equation}
t=t_S +2 \sqrt{2 m r} +2 m \ln \frac{\sqrt{r}-\sqrt{2 m}}{\sqrt{r}+\sqrt{2 m}}.
\end{equation}
The Schwarzschild metric can then be written as
\begin{equation}
\label{pan}
ds^2=\left(1-\frac{2 m}{r}\right) dt^2 - 2 \sqrt{\frac{2 m}{r}} dt
dr-dr^2-r^2 d\Omega^2.
\end{equation}
When considering events occurring at $r=r_{2m} \equiv 2 m$ this coordinate system has
a number of advantages over Schwarzschild coordinates.  Firstly,
surfaces of constant $t$ cross the surface $r=r_{2m}$, allowing events
along $r_{2m}$ to be assigned a time at which they occur.
Secondly, as mentioned above, $t$ corresponds to the proper time of a
freely falling observer along a radial geodesic parameterised by
$\dot{r} = -\sqrt{2 m/r}$ (over-dots denote differentiation with respect to $t$ throughout).
Furthermore, surfaces of constant $t$ are simply Euclidean 3-spaces.

To account for the reduction in mass of the hole due to
tunnelling we now parameterise $m$ as $m(t)$.
(If one were to attempt such a parameterisation in terms of $t_S$ then
a curvature singularity would appear at $r_{2 m}$).  This time-dependent
Panlev\'{e} geometry is the one considered by Brown and Lindesay in
\cite{lind1, lind2, lind3}.  We consider it to be the continuum limit
of the near horizon geometry that results from many small, discrete tunnelling
events.  For simplicity we
consider radiated particles to be s-waves, so that by Birkhoff's
theorem their effects on the geometry of the hole can be neglected, other than
by reducing the value of $m$.  This geometry is similar to the
Vaidya metric, but here the mass, $m$, is made a function of the
Panlev\'{e} time, $t$, used in the tunneling calculation, instead of
the advanced time, $v_s=t_s+r$.  This geometry allows the mass loss due to
tunneling to be taken into account straightforwardly, as
well as allowing the metric to be cast in double null co-ordinates in a
simple way.

Consider the retarded and advanced null coordinates \cite{lind2}
\begin{align}
\label{u1}
du &= A \left(1-\sqrt{\frac{2 m}{r}}\right) dt - A dr\\
dv &= B \left(1+\sqrt{\frac{2 m}{r}}\right) dt + B dr,
\label{v1}
\end{align}
where $A=A(t,r)$ and $B=B(t,r)$ must satisfy the integrability conditions
\begin{align}
\label{A1}
\frac{\partial}{\partial r} \left( \left( 1-\sqrt{\frac{2 m}{r}}
\right) A\right)+ \frac{\partial A}{\partial t} &=
0\\
\label{B1}
\frac{\partial}{\partial r}\left( \left( 1+\sqrt{\frac{2 m}{r}}
\right) B\right)-\frac{\partial B}{\partial t} &=
0.
\end{align}
If $m=$constant then these equations have the simple solutions $A= 1/(1-\sqrt{2 m/r})$ and
$B= 1/(1+\sqrt{2 m/r})$, corresponding to the Schwarzschild solution
in null coordinates.  If $m=m(t)$ then for a slowly evaporating black hole we can
approximate $m$ by an expansion about some early moment, $t_0$, so that
$m(t) \simeq m(t_0)+\dot{m}(t_0) t$.  To this order of approximation
equations (\ref{A1}) and (\ref{B1}) have the solutions
\begin{align}
\label{A2}
A &= \prod_i \left( x_i-\sqrt{\frac{2 m}{r}} \right)^{\frac{x_i}{2-3 x_i}}\\
\label{B2}
B &= \prod_i \left( y_i+\sqrt{\frac{2 m}{r}} \right)^{\frac{y_i}{2-3 y_i}}
\end{align}
where $x_i$ are the roots of $2 \dot{m}-x^2+x^3=0$, and $y_i$ are the
roots of $2 \dot{m} +y^2-y^3=0$.  In terms of the new coordinates,
defined by (\ref{u1}) and (\ref{v1}), the line-element (\ref{pan}) then takes
the form
\begin{equation}
ds^2 =\frac{du dv}{A B}-r^2 d\Omega^2,
\end{equation}
where $A$ and $B$ are given by (\ref{A2}) and (\ref{B2}).  This geometry will
be used to describe the region outside a shell of matter that
collapses to form a radiating black hole.  Inside, the geometry will be
taken to be Minkowski space covered by the retarded and advanced null
coordinates, $U=T-r$ and $V=T+r$, such that
$ds^2=dUdV-r^2 d\Omega^2$.
The two sets of null coordinates in the two regions can then be related at the
boundary between them by the transformations $\alpha(u)=U$ and
$\beta(V)=v$.  If this boundary is at a radial distance $r=R(t)$, then
these functions are given by the differential relations
\begin{align}
\label{alpha}
\alpha^\prime &= \frac{dU}{du} = \frac{\sqrt{1-\frac{2
      m}{R}-2\sqrt{\frac{2 m}{R}} \dot{R}}-\dot{R}}{A \left(
      1-\sqrt{\frac{2 m}{R}} -\dot{R}\right)}\\
\beta^\prime &= \frac{dv}{dV} =\frac{B \left( 1+\sqrt{\frac{2 m}{R}} +\dot{R} \right)}{\sqrt{1-\frac{2
      m}{R}-2\sqrt{\frac{2 m}{R}} \dot{R}}+\dot{R}}.
\label{beta}
\end{align}

\section{Tunnelling and the horizon}

\begin{figure}[ht]
\centering
\epsfig{figure=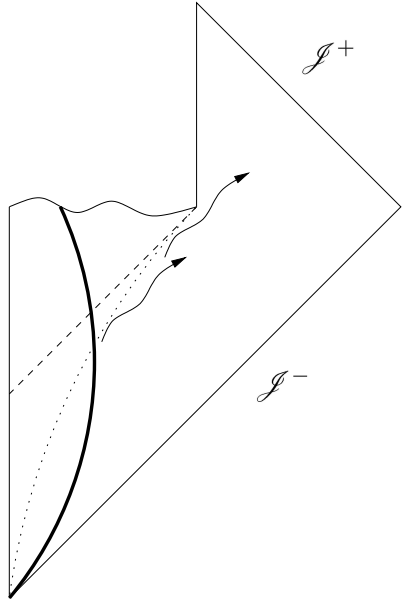, height=8cm}
\caption{The Penrose diagram of the geometry (\ref{pan}).  The dashed
  line is the global horizon $r_H$, the dotted line is the surface
  $r=2m$, and the thick solid line is the trajectory of the shell,
  $R(t)$.  Radiation is shown being emitted from $r_{2m}$ to future null
  infinity, $\mathscr{J}^+$.}
\label{pen}
\end{figure}

An important feature of the geometry (\ref{pan}) is that the global
event horizon (the boundary of the past of future null infinity) is
\textit{not} at $r=2m$.  It is displaced by a small
amount\footnote{Although it has been suggested by some that it may be displaced by a
large amount, so that an event horizon may not form at all
\cite{vach1, niel2}.} \cite{lind1,
  lind2,lind3}.  As the global event horizon
is a null surface it is described by the outgoing null geodesics
\begin{equation}
\dot{r}_H = 1-\sqrt{\frac{r_{2m}}{r_H}},
\end{equation}
so that $r_H = r_{2m}$ only when $\dot{r}_H \rightarrow 0$, the limit
where the black hole
is not evaporating.  For an evaporating hole we have $\dot{r}_H<0$, and so
$r_H<r_{2m}$.  Under the approximation $m(t)= m(t_0)+\dot{m}(t_0) t$
the location of $r_H$ can then be seen to be given by the positive
real solution of
\begin{equation}
2 \dot{m}-\frac{r_{2m}}{r_H}+\left(\frac{r_{2m}}{r_H}\right)^{3/2}=0.
\end{equation}
This is exactly the point at which $A$ becomes singular, as can be
seen from (\ref{A2}).  The Penrose diagram for this geometry (see
\cite{lind2} for details) is shown
in figure \ref{pen}, where the global horizon at $r_H$ is represented
by a dashed line and the surface $r=2m$ is given by a dotted line.
The thick solid line represents the trajectory of the shell, $R(t)$.

The question now arises, through which surface are
the radiated particles tunnelling: $r=r_{2m}$ or $r=r_H$?  To answer this
let us consider the rate of tunnelling $\Gamma \sim \exp\{-2 \text{Im
    S}\}$, where S is the action along the classically forbidden
trajectory.  Using Hamilton's equation, $\dot{r}=dH/dp_r \vert_r$,
Parikh and Wilczek \cite{par1} find the exponent of this rate to be
\begin{equation}
-2 \text{Im S}= -2 \text{Im} \int_{r_{in}}^{r_{out}} p_r dr= -2 \text{Im}
 \int_{r_{in}}^{r_{out}} \int_{0}^{p_r}dp_r^\prime dr = -2\text{Im}
 \int \int \frac{dr}{\dot{r}} dH.
\end{equation}
Here $p_r$ is the radial momentum of the particle that is tunnelling
from $r_{in}$ to $r_{out}$.  The particles trajectory $\dot{r}$ is
then given by the null geodesics of (\ref{pan}) with $m$ replaced
by $m-\omega$, where $\omega$ is the energy of the particle.  They then
swap the order of integration and obtain
\begin{equation}
-2 \text{Im S}= -2 \text{Im} \int_0^{\omega} \int_{r_{in}}^{r_{out}}
 \frac{dr}{1-\sqrt{\frac{2 (m-\omega^{\prime})}{r}}} (-d\omega^\prime)
=-8\pi \omega\left(m-\frac{\omega}{2}\right)
\end{equation}
where the final expression is given provided $r_{in}>r_{out}$.  It is
then noted that if they had instead kept the original order of
integration they would have found
\begin{equation}
-2 \text{Im S} = -2 \text{Im} \int_{r_{in}}^{r_{out}}
 \int_{m}^{m-\omega} \frac{dm^{\prime}}{1-\sqrt{\frac{2 m^\prime}{r}}}
 dr = 2 \pi \int_{r_{in}}^{r_{out}} r dr.
\end{equation}
If $r_{out}= r_{in}-2 \omega$, the value of $r_{in}$ is then $2 m$.  This shows
that the tunnelling events occur through the
surface $r_{2m}$, where the action obtains an imaginary part.  This is
distinct from the global event horizon at $r=r_H$.
For further details the reader is referred to the original text
\cite{par1}, and for criticisms and discussion to
\cite{con}.  This result has also be generalised to other cases \cite{others}.

The idea that radiation should be emitted from a surface outside
the horizon is not new:  It has been considered by Visser
\cite{visser}, di Criscienzo \textit{et al} \cite{di}, and Nielsen
\cite{niel1}, as well by Susskind, Thorlacius and Uglum \cite{stretch} in the form
of a `stretched horizon'.  The stretched horizon was envisaged as a place where
information about the micro-physical structure of objects falling into
a black hole could be stored, and subsequently encoded in
outgoing radiation.  In the present study, as in \cite{visser, di, niel1}, this surface is the place where an outgoing
particle trajectory is classically forbidden.  Such a surface is a physically
meaningful concept at finite times, unlike the surface $r=r_H$ which
requires a knowledge of the global structure of space-time in
order to be defined.

\section{The energy-momentum of radiation}

We will now compute the vacuum expectation value of the
energy-momentum tensor of a massless conformal scalar
field in the vicinity of the surface $r_{2m}$, outside of a collapsing
shell of matter.  This calculation will be in two dimensions and will follow the study of Davies
\cite{dav1}, which was performed with $m=$constant.  The
regularisation method employed will be the point splitting method of
Davies and Fulling \cite{dav2}.  In this method the product of field
operators in the energy-momentum tensor are evaluated as functions of two points along
a geodesic.  The points are then brought together, and the divergent
terms covariantly subtracted.  The result is a non-divergent and covariant,
renormalised vacuum expectation value for the energy-momentum tensor.
Although the two dimensional case has proved a useful tool in
understanding semi-classical radiation, one should keep in mind that it
may not be exactly analogous to the corresponding case in four
dimensions \cite{fro}.  This method has been previously applied to
dynamical black hole space-times in \cite{his, bal}.

Following \cite{dav3} we introduce a new system of null coordinates,
$\bar{u}$ and $\bar{v}$, that cover the space-time both inside and
outside the shell of radius $R(t)$.  The metric is then given by
\begin{equation}
\label{null2}
ds^2=C(\bar{u},\bar{v}) d\bar{u}d\bar{v}
\end{equation}
where
\begin{eqnarray*}
C(\bar{u},\bar{v}) =&\frac{dU}{d\bar{u}} \frac{dV}{d\bar{v}} \qquad &r<R(t)\\
=&\frac{1}{AB} \frac{du}{d\bar{u}} \frac{dv}{d\bar{v}} \qquad &r>R(t).
\end{eqnarray*}
As this metric is conformally flat, the solutions of the scalar wave
equation are the solutions of
$\phi_{,\bar{u} \bar{v}} = 0$,
which are simply $e^{-i \omega \bar{v}}$ and $e^{-i \omega
\bar{u}}$.  The choice $\bar{v}=v$ ensures these fields can be made to correspond with their usual
form in Minkowski space at past null infinity.  Furthermore,
choosing $\bar{u}=\beta(U)$ means that outgoing waves $e^{-i \omega
  \bar{u}}$ can be matched to incoming waves $e^{-i \omega v}$ at $r=0$,
allowing a simple correspondence as they pass through the
centre of the shell.

The point splitting procedure of \cite{dav2} now yields the
renormalised energy-momentum tensor for a massless conformal field in
the geometry (\ref{null2}) as
\begin{equation}
T_{\bar{\mu} \bar{\nu}} = \theta_{\bar{\mu} \bar{\nu}} +\frac{R}{48
  \pi} g_{\bar{\mu} \bar{\nu}}
\end{equation}
where $\theta_{\bar{u} \bar{u}} = -F_{\bar{u}}(C)$, $\theta_{\bar{v}
  \bar{v}} = -F_{\bar{v}}(C)$, $\theta_{\bar{u} \bar{v}} =
  \theta_{\bar{v} \bar{u}} = 0$
and $F_x(y)=(12 \pi)^{-1} y^{1/2} (y^{-1/2})_{,x x}$.
The components of this tensor can be evaluated using the relations
specified above between the three coordinate systems.  Inside the
shell the non-zero components are, in $U$ and $V$ coordinates,
\begin{align}
\label{1}
T_{UU} &= F_U(\beta^\prime(U))\\
\label{2}
T_{VV} &= F_V(\beta^\prime(V)),
\end{align}
whilst outside the shell we have, in $u$ and $v$ coordinates,
\begin{align}
\label{3}
T_{uu} &= \frac{1}{3A^2} F_r\left(e^{-3\sqrt{\frac{2 m}{r}}}\right) +
\frac{1}{A^2} F_r(A) +\sqrt{\frac{2 m}{r}} \frac{\dot{m}}{192 \pi r m
  A^2} +{\alpha^\prime}^2(u) F_U(\beta^\prime(U))
+F_u(\alpha^\prime(u))\\
\label{4}
T_{vv} &= \frac{1}{3 B^2} F_r \left(e^{3\sqrt{\frac{2m}{r}}}\right)
+\frac{1}{B^2} F_r(B)+\sqrt{\frac{2m}{r}} \frac{\dot{m}}{192 \pi r m
  B^2}\\
\label{5}
T_{uv} &= T_{vu} = -\frac{1}{192 \pi rAB} \left(
\frac{8m}{r^2}+\sqrt{\frac{2m}{r}}\frac{\dot{m}}{{m}}\right),
\end{align}
where $A$, $B$, $\alpha^\prime$ and $\beta^\prime$ are
given by (\ref{A2}), (\ref{B2}), (\ref{alpha}) and
(\ref{beta}).  The expressions (\ref{1})-(\ref{5}) now give the required form of the
energy-momentum tensor for a scalar field in the presence of a
collapsing shell of matter, and will be used below to investigate the
energy density of radiation measured by observers
watching the collapse.

\section{Measurements made by asymptotic observers}

\subsection{$\mathbf{m=\text{constant}}$}

Let us briefly review the static limit, $m\rightarrow
\text{constant}$, so that $A\rightarrow 1/(1-\sqrt{2 m/r})$, $B\rightarrow 1/(1+\sqrt{2
  m/r})$ and the results of \cite{dav1} are retrieved.  In this case the third
terms  in (\ref{3}) and (\ref{4}), and the second term in
(\ref{5}), are zero everywhere.  The first two terms in each of
(\ref{3}) and (\ref{4}) are identical, and, along with the first
term of (\ref{5}), represent an energy density in the vicinity of
the black hole due to the curvature of Schwarzschild space-time.
This energy density vanishes in the limit $r\rightarrow \infty$, and
does not contribute to the Hawking radiation measured by asymptotic
observers.

The remaining two terms of (\ref{3}) are functions of the
retarded time coordinates only, and can therefore give a non-zero energy
density at future null infinity.  The last of these terms,
$F_u(\alpha^\prime(u))$, is a function of $R(t)$ for finite $u$.  As $u\rightarrow
\infty$, however, it approaches $1/768 \pi m^2$ and so corresponds to the Hawking term.  The penultimate term
in (\ref{3}), ${\alpha^\prime}^2(u) F_U(\beta^\prime(U))$, matches the
mode inside the shell, given by equations (\ref{1}) and (\ref{2}), which is initially
in-going, then passes through the centre at $r=0$ before becoming an outgoing ray
that passes back out through the shell and on to
future null infinity.  This term is sensitive to the form of
$R(t)$, and its pre-factor of ${\alpha^\prime}^2$,
due to the transformation from $U$ to $u$, is non-zero for finite $u$.
As $u \rightarrow \infty$, however, $\alpha^\prime
\rightarrow 0$ so that the contribution of this term vanishes in that limit.

When $m=\text{constant}$ the only relevant non-zero term as $u
\rightarrow \infty$ is therefore
the Hawking one, which is a constant, independent of the process that
formed the black hole (i.e. independent of $R(t)$).  For finite $u$
this is not the case, and the flux of radiation reaching future null
infinity is sensitive to the collapse process.  Whether this
corresponds to any useful information reaching the asymptotic
observer is unclear, as in this case surfaces of finite $u$ can be
removed to the distant past by the stacking of surfaces of constant
$u$ against the event horizon.  We will now
investigate the case in which $\dot{m} \neq 0$.  As discussed above
the source of the radiation in this case is expected to be the surface
$r=r_{2m}$, which is displaced from the global event horizon at $r_H$.

\subsection{$\mathbf{m\neq \text{constant}}$}

It can be seen that when $\dot{m} \neq 0$ the first two terms of
(\ref{3}) and (\ref{4}), and the first term of (\ref{5}), are still
independent of the collapse process.  Their forms are modified from the
$\dot{m} = 0$ case, as the two functions $A$ and $B$ are now
given by (\ref{A2}) and (\ref{B2}), but they still approach zero as
$r\rightarrow \infty$.  When $\dot{m} \neq 0$ we now have three new
terms:  The third terms in (\ref{3}) and (\ref{4}), and the second term in
(\ref{5}).  These terms also vanish as $r\rightarrow \infty$, and so
do not contribute to any flux at future null infinity\footnote{These terms may be badly behaved
  in the limit $m\rightarrow 0$ due to an unrealistic aspect of the
  model.  This will be commented upon in the discussion section.  For further discussion of diverging
  energy-momentum in dynamical black hole space-times see \cite{his, singh}.}.

The only non-zero components of the energy-momentum tensor, as $r
\rightarrow \infty$, are therefore
\begin{equation*}
T_{uu} \rightarrow {\alpha^\prime}^2(u) F_U(\beta^\prime(U))
+F_u(\alpha^\prime(u)).
\end{equation*}
These are the same two terms that contribute to the asymptotic flux in the $\dot{m}=0$ case,
but now the $\alpha$ and $\beta$ terms, as well as the coordinate
transformations between $u$, $U$ and $t$, are modified by the
non-constant $m(t)$.  Substituting (\ref{A2}), (\ref{B2}), (\ref{alpha})
and (\ref{beta}) into the expression above, together with the relevant
transformations from $u$ and $U$ to $t$, allows these two terms to
be evaluated as functions of $t$.  Due to the complicated
form of the resulting expressions we choose not to show them
explicitly here, but rather to present them as plots.  This will
demonstrate how the present case differs from the $m=\text{constant}$
case, which we will display in parallel for comparison.

\begin{figure}[htbp]
\centering
\epsfig{figure=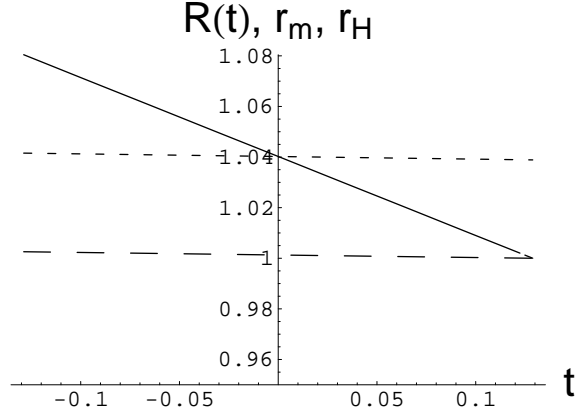, height=5.5cm}
\caption{The trajectories of $R$, $r_{2m}$ and $r_H$ when $m= 1-0.01
t$ and $R= 2-0.6 t$, shown as a solid line, a short-dashed line
and a long-dashed line, respectively.  The shell crosses $r_{2m}$ at
$t=0$, and the global horizon at $t\simeq 0.13$.}
\label{plot1}
\end{figure}

For illustrative purposes let us consider $m = 1-0.01
t$ and $R = 2-0.6 t$, so that $R=r_{2m}$ at $t=0$.  The trajectories of $R$, $r_{2m}$ and $r_H$
are then shown in figure \ref{plot1} as the solid line, the dotted line
and the dashed line, respectively.  Before $t=0$ the shell is
outside of both $r_{2m}$ and $r_H$.  Between $t=0$ and $t\simeq 0.13$ the shell is
between $r_{2m}$ and $r_H$, and after $t \simeq 0.13$ the shell is inside the
global horizon.  We will refer to this choice of parameters as ``hole 1''.

\begin{figure}[htbp]
\center
\subfigure[$F_u(\alpha^\prime (u))$ for the three holes, as a function
  of $t$ along $R(t)$.]{\epsfig{figure=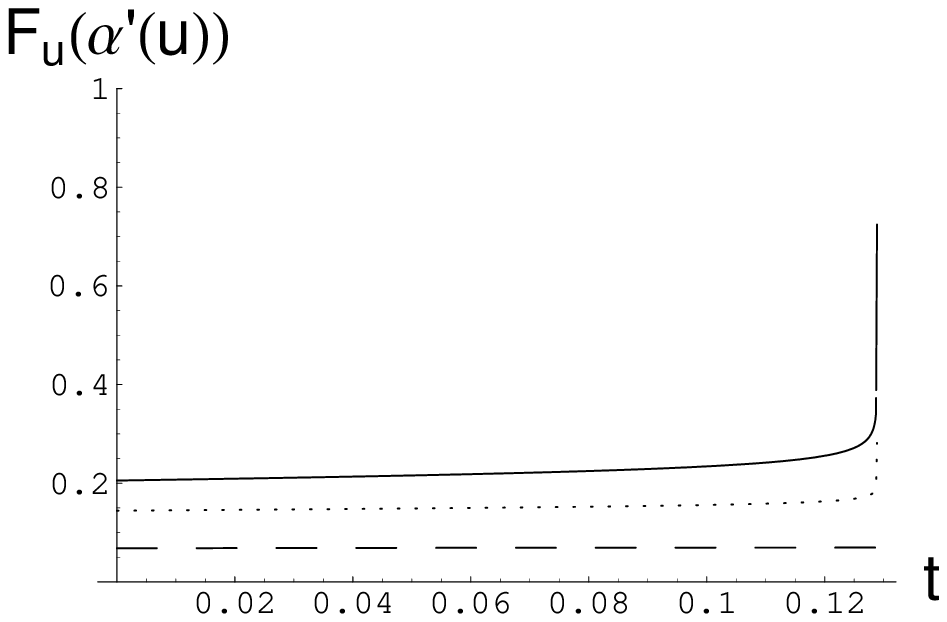,height=4.9cm}}
\qquad
\subfigure[${\alpha^\prime}^2(u) F_U(\beta^\prime (U))$ for the three
  holes, as a function
  of $t$ along $R(t)$.]{\epsfig{figure=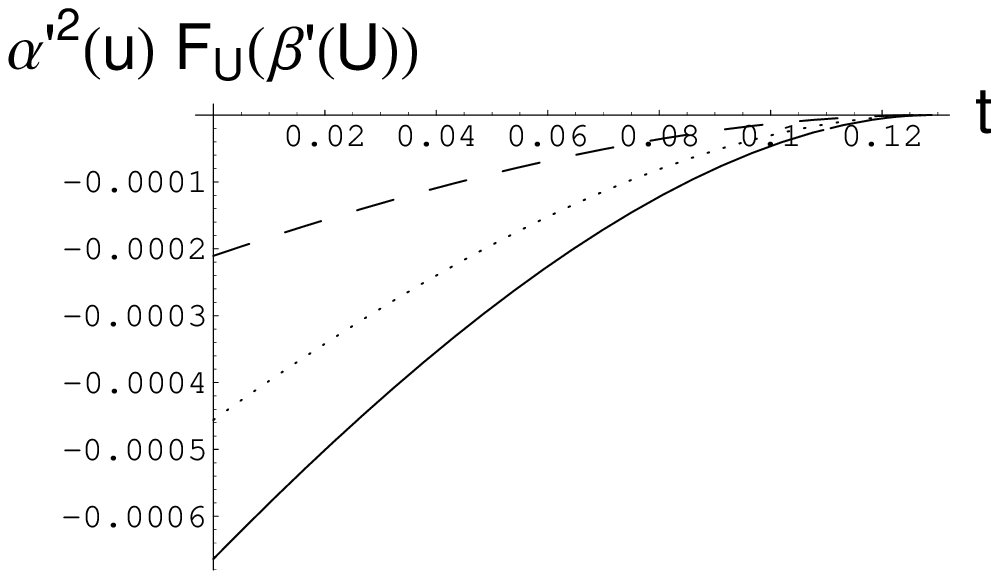,height=4.9cm}}
\caption{The two contributions to the flux of
  radiation at future null infinity for hole 1 (the solid line), hole
  2 (the dotted line) and hole 3 (the dashed line).}
\label{plot23}
\end{figure}

$F_u(\alpha^\prime (u))$ for hole 1 can be readily calculated and is
shown in figure \ref{plot23}(a) as a function of $t$ along the trajectory
$R(t)$.  Similarly, ${\alpha^\prime}^2(u) F_U(\beta^\prime (U))$
is shown in figure \ref{plot23}(b).  These plots have been normalised
so that $F_u(\alpha^\prime (u))=1$ when $R=r_H$, the moment the shell
crosses the global event horizon.  For comparison we also show in figures \ref{plot23}(a)
and \ref{plot23}(b) the results that would have been obtained for a
black hole evaporating at half the rate (``hole 2'', the dotted lines) and a black
hole with constant mass (``hole 3'', the dashed lines).  In both of
these cases the trajectory of the in-falling shell is kept the same as
for hole 1, and both have been chosen so that the $r_H$ of all three
holes are equal at the moment $R=r_H$.
It can be seen from figure \ref{plot23} that the value of
$F_u(\alpha^\prime (u))$ for hole 3 is approximately constant over the
time scale being considered, at the Hawking rate of $1/768\pi m^2$.
For holes 1 and 2 $F_u(\alpha^\prime (u))$ can be seen to be a slowly
increasing function of $t$, that becomes rapidly increasing as $R
\rightarrow r_H$.  The value of ${\alpha^\prime}^2(u) F_U(\beta^\prime
(U))$ can be seen to approach zero for all three holes as
$R\rightarrow r_H$, as expected.  This is due to ${\alpha^\prime}(u)
\rightarrow 0$ in this limit.

\begin{figure}[htbp]
\center
\subfigure[$F_u(\alpha^\prime (u))$ as a function
  of $t_{\infty}$.]{\epsfig{figure=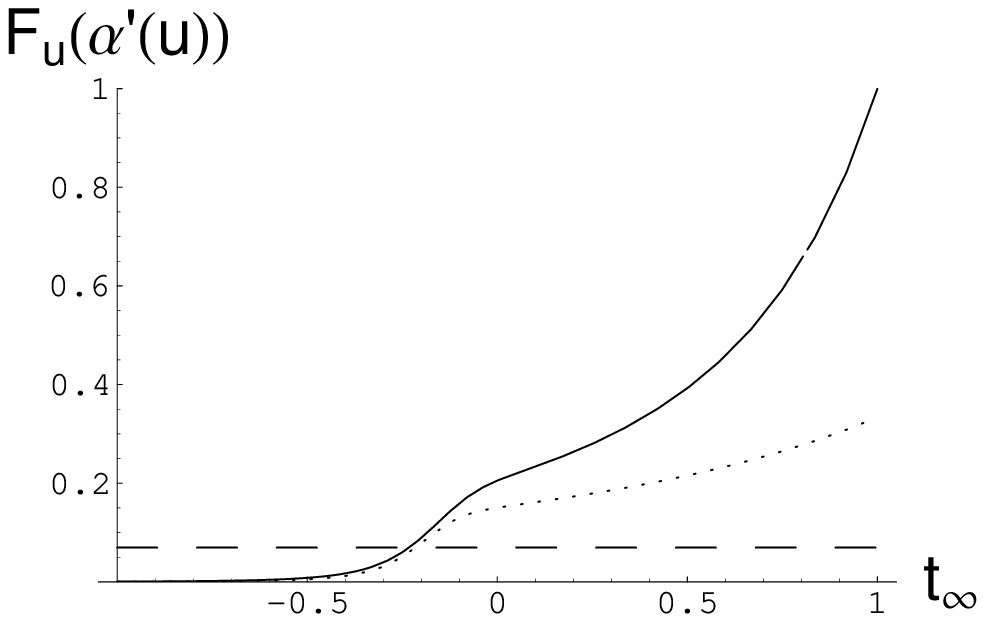,height=4.9cm}}
\qquad
\subfigure[${\alpha^\prime}^2(u) F_U(\beta^\prime (U))$ as a function
  of $t_{\infty}$.]{\epsfig{figure=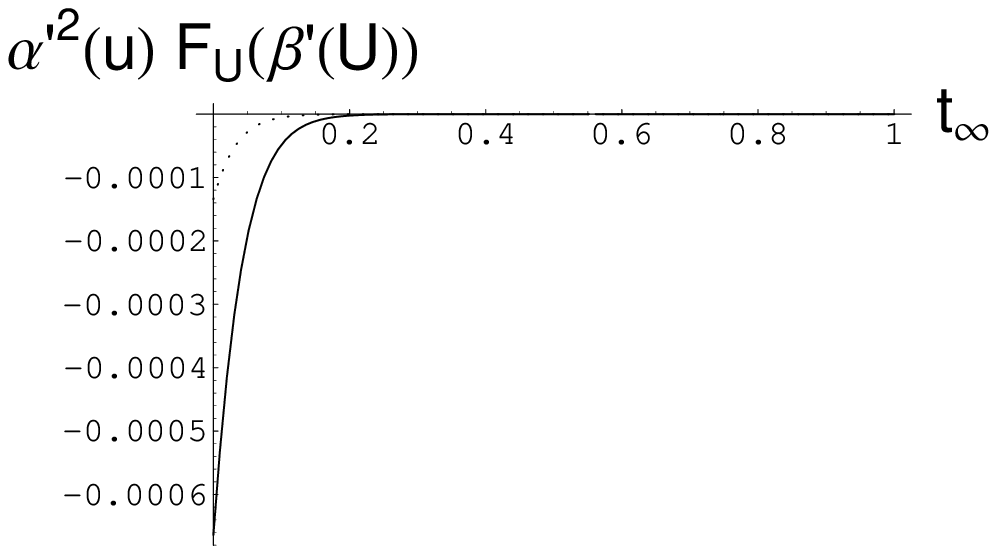,height=4.9cm}}
\caption{The two contributions to the flux of
  radiation at future null infinity for hole 1 (the solid line), hole
  2 (the dotted line) and hole 3 (the dashed line), as measured in the proper
  time of an observer there.}
\label{plot45}
\end{figure}

Let us now consider what the
observer at infinity measures.  To find this we must transform the plots above
from $t$ to $t_{\infty}$, the proper time of an asymptotic observer, by
\begin{equation}
\label{tinf}
dt_{\infty} = \left[ A(t) \left( 1-\sqrt{\frac{2
      m(t)}{R(t)}}-\dot{R}(t)\right) \right] dt.
\end{equation}
This expression relates the two time coordinates via the retarded time $u$.  The
two terms $F_u(\alpha^\prime (u))$ and  ${\alpha^\prime}^2(u) F_U(\beta^\prime (U))$ are
now given as function of $t_{\infty}$ in figures \ref{plot45}(a) and \ref{plot45}(b).  These
plots are normalised by the value of $F_u(\alpha^\prime (u))$ for
hole 1 at $R=r_H$, and the new time coordinate for holes 1 and 2 have
been chosen so that $t_{\infty}=0$ when the asymptotic observer sees
$R=r_{2m}$.  (This is not possible in the static case as the surface
$r_{2m}$ is degenerate with $r_H$).  Units have been chosen so that $t_{\infty}=1$ when the
asymptotic observer sees $R=r_H$, in each case.
These plots are the same as those of figure \ref{plot23}, but they have been
stretched near $R=r_H$ and squashed near $R=r_{2m}$, giving them a considerably different
appearance.  The stretching involved increases as the rate of
evaporation decreases, diverging in the limit of a static hole.  Hence
the lines corresponding to hole 3 are now straight, and equal to zero in the
case of figure \ref{plot45}(b).

The observer at infinity watching holes 1 and 2 sees an energy density
in the mode $F_u(\alpha^\prime (u))$ that rises as the
shell is seen to pass $r_{2m}$, and subsequently increases at a steady
rate.  This is due to the surfaces of constant $u$ stacking up against
the global horizon, and stretching out the spikes seen in figure
\ref{plot23}(a).  That the radiation increases as the shell crosses
$r_{2m}$ is in support of the conjecture that the radiation is originating
from this surface, as found directly in the tunnelling calculation.

Figure \ref{plot45}(b) shows that for holes 1 and 2
the contribution from ${\alpha^\prime}^2(u) F_U(\beta^\prime (U))$, as
witnessed by the asymptotic observer, becomes exponentially small soon
after the shell crosses $r_{2m}$.  Again, this is
due to the stacking of surfaces of constant $u$ against the global
horizon, and the resulting stretching of figure \ref{plot23}(b).  In
the vicinity of $R=r_{2m}$, however, while this component is small, it
is still non-zero.  This is of interest as $F_U(\beta^\prime (U))$ is
known to be sensitive to the collapse process that formed the hole \cite{dav1}.

\begin{figure}[htbp]
\center
\subfigure[$F_u(\alpha^\prime (u))$ as a function
  of $t$.]{\epsfig{figure=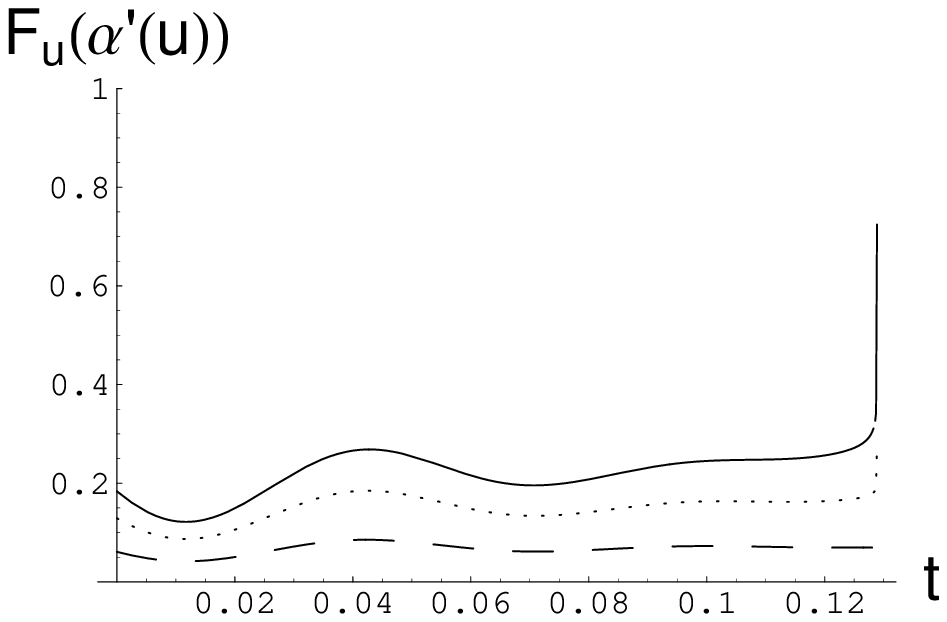,height=4.9cm}}
\qquad
\subfigure[${\alpha^\prime}^2(u) F_U(\beta^\prime (U))$ as a function
  of $t$.]{\epsfig{figure=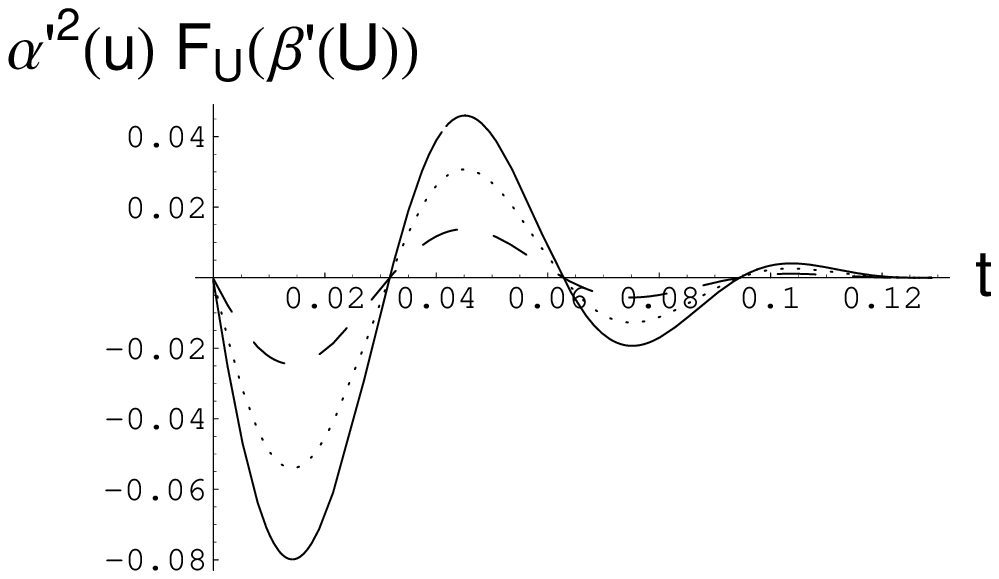,height=4.9cm}}
\caption{The contributions to the flux of
  radiation at future null infinity for hole 1 (the solid line), hole
  2 (the dotted line) and hole 3 (the dashed line), for the
  perturbed shell.}
\label{plot78}
\end{figure}

To consider the degree to which the outgoing radiation is sensitive to
the collapse process, let us now perturb the in-falling trajectory of the
shell by a small amount, so that $R = 3 \times 10^{-5} \cos (10^2 t)+ 2-0.6
t$.  This is a small enough perturbation that the plot of the
trajectory of the in-falling shell is indistinguishable from figure
\ref{plot1}.  However, the effects on $F_u(\alpha^\prime (u))$ and
${\alpha^\prime}^2(u) F_U(\beta^\prime (U))$ are not negligible, and
are shown in figure \ref{plot78}.  The same scalings and origin have
been chosen as above.  Perturbations to the infall trajectory could result from imperfections in spherical
symmetry, internal stresses, or interactions of the
shell with other objects.  Here we simply wish to illustrate
the sensitivity of the emitted raditiation to the collapse process, and so add an {\it ad hoc} oscillation
to the shell's trajectory.
It can be seen such oscillations, although small, have a discernible effect on the
outgoing radiation of the holes.  Both functions $F_u(\alpha^\prime (u))$ and
${\alpha^\prime}^2(u) F_U(\beta^\prime (U))$ can be seen to oscillate around
their unperturbed values, as shown in figure \ref{plot23}.  These
oscillations are particularly apparent in ${\alpha^\prime}^2(u)
F_U(\beta^\prime (U))$.

\begin{figure}[htbp]
\center
\subfigure[$F_u(\alpha^\prime (u))$ as a function
  of $t_{\infty}$.]{\epsfig{figure=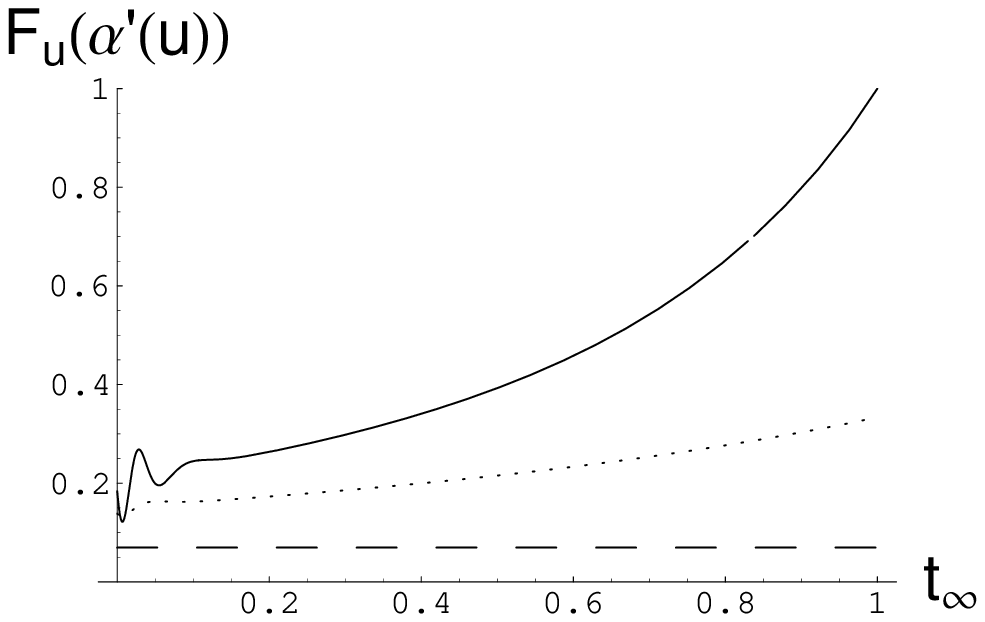,height=4.9cm}}
\qquad
\subfigure[${\alpha^\prime}^2(u) F_U(\beta^\prime (U))$ as a function
  of $t_{\infty}$.]{\epsfig{figure=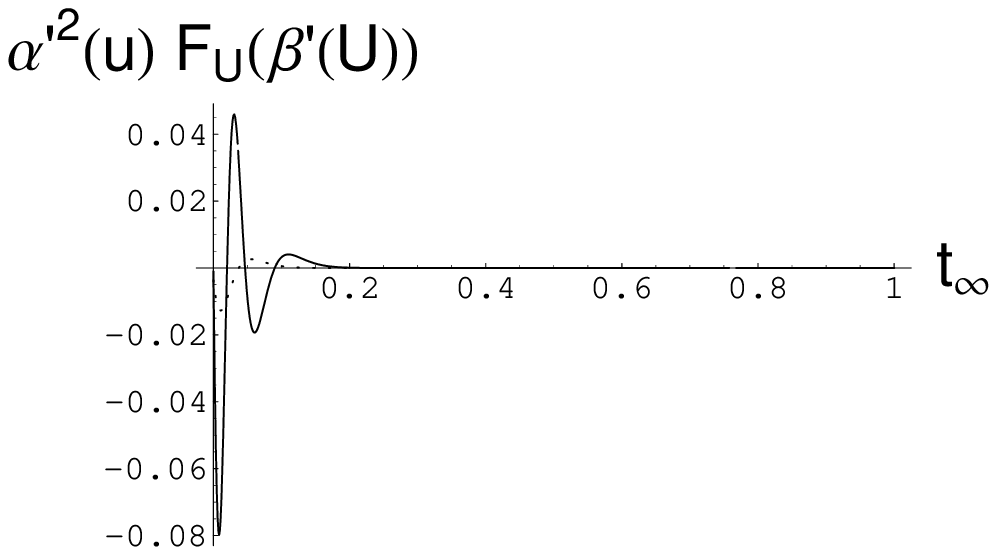,height=4.9cm}}
\caption{The contributions to the flux of
  radiation at future null infinity from hole 1 (the solid line), hole
  2 (the dotted line) and hole 3 (the dashed line) in terms of $t_{\infty}$, for the
  perturbed shell.}
\label{plot910}
\end{figure}

Now let us consider the radiation measured by an asymptotic observer
watching the perturbed shell collapse.  Any features present in the
plots in figure \ref{plot78} should also be present in the radiation he/she measures, although
it will be effected by the stretching and squashing involved in the
transformation from $t$ to $t_{\infty}$.  Using (\ref{tinf}) we find
that $F_u(\alpha^\prime (u))$ and ${\alpha^\prime}^2(u)
F_U(\beta^\prime (U))$  are given as functions of $t_{\infty}$ as shown
in figure \ref{plot910}.  For holes 1 and 2, the oscillations that were present in
figure \ref{plot78} are again apparent, but now restricted
to be observable only in the vicinity of $R=r_{2m}$ (due to the
stacking of surfaces of constant $u$ against the global horizon).  The infinite stretching in
the case of hole 3 is enough to remove the oscillations that were
present in figure \ref{plot78} to the infinite past.

\section{Discussion}

We have considered the case of a black hole that forms from the
collapse of a spherical shell, and decays via tunnelling events.  In
this case the mass parameter of the hole is made a function of
the Panlev\'{e} time, $t$, and the surface $r=2 m$ separates
from the global event horizon by a small amount.  The tunnelling events occur
at $r=2m$, and not at the global event horizon.  This displacement has consequences for
the emission of Hawking radiation, and for its observation by asymptotic observers
watching the collapse.

The energy-momentum tensor of a massless
conformal scalar field is obtained in the two dimensional analogue
of this space-time, and the flux of radiation measured by an
asymptotic observer is calculated.    
The asymptotic observer witnesses radiation that initially
increases at about the time he/she sees the collapsing shell cross $r=2m$, and
increases at a steady rate thereafter.  
This supports the idea that the
Hawking radiation is emitted from the vicinity of the surface $r=2m$
\cite{visser,di,niel1},
and that after this surface is exposed to asymptotic observers he/she witnesses radiation from a hole
that is slowly increasing in temperature.

Radiation originating from outside of
the global event horizon contains modes that are sensitive to the
collapse process \cite{dav1}.  This sensitivity is illustrated
by perturbing the collapsing shell with a small oscillation.  The effects of this on the
modes that escape to future null infinity are clearly visible, and
only vanish in the limit that the global horizon is approached.  For
the static model the infinite stacking of surfaces of constant retarded time
against the global horizon removes these modes to the remote past, so
that it is unclear whether they have any observational significance.
This is not the case in the non-static models.
Here the sensitivity of the emitted radiation to the collapse process
is clearly visible, but restricted to be observable only at the
beginning of the evaporation process.  Nevertheless,
an observer who measures all of the radiation emitted from
the hole may, in principle, be able to extract some
information about the collapse process that formed it.

There are a number of ways in which this study could be improved.  In
particular, one could parameterise $m$ as a function of
$t$ in a more satisfactory way.  Here we assumed a function of the form $m=m_0+\dot{m}_0 t$,
in order to get explicit, exact solutions.  While this is likely to be valid
for the early stages of evaporation (being the first term of a
series expansion), it is unlikely to be a good approximation
during the late stages.  Ideally one would like to find solutions with
more general $m(t)$.  A further problem is the behaviour
of the model as $m\rightarrow 0$, when the black hole evaporates
completely \cite{his,singh}.  In this limit (depending on the form of $m(t)$) the
scalar curvature invariants associated with the space-time (\ref{pan}) may diverge, indicating a
singularity\footnote{This is the origin of the bad behaviour mentioned
in the footnote of section 5.2.}.  This shows that a
continuum limit for $m(t)$ is not appropriate at the end point of
evaporation, and that a cutoff should be introduced beyond which a
more appropriate method is used.
\\\\
\leftline{\bf Acknowledgements}

I am grateful to Pedro Ferreira and Philip Candelas for helpful
discussions, and to Jesus College for support.


\begin{thebibliography}{100}

\bibitem{hawk1} S. W. Hawking, Comm. Math. Phys. \textbf{43}, 199 (1975).

\bibitem{par1} M. K. Parikh and F. Wilczek, Phys. Rev. Lett.
  \textbf{85}, 5042 (2000).

\bibitem{kraus} P. Kraus and F. Wilczek, Nucl. Phys. B \textbf{433},
  403 (1995). E. Keski-Vakkuri and P. Kraus, Nucl. Phys. B
  \textbf{491}, 249 (1997).

\bibitem{lind1} J. Lindesay, Foundations of Physics \textbf{37}, 1181
  (2007). 

\bibitem{lind2} B. A. Brown and J. Lindesay, arXiv:0710.2032 (2007).  

\bibitem{lind3} B. A. Brown and J. Lindesay, arXiv:0802.1660 (2008).

\bibitem{vach1} T. Vachaspati, D. Stojkovic and L. M. Krauss,
  Phys. Rev. D \textbf{76}, 024005 (2007).  T. Vachaspati and D. Stojkovic, Phys. Lett. B
  \textbf{663}, 107 (2008).

\bibitem{dav1} P. C. W. Davies, Proc. R. Soc. Lond. \textbf{351}, 129 (1976).

\bibitem{dav2} P. C. W. Davies and S. A. Fulling,
  Proc. R. Soc. Lond. \textbf{348}, 393 (1975).

\bibitem{his} W. A. Hiscock, Phys. Rev. D \textbf{23}, 2813 (1981).

\bibitem{bal} R. Balbinot, Phys. Rev. D \textbf{33}, 1611 (1986).

\bibitem{pan} P. Panlev\'{e}, C. R. Acad. Sci. (Paris) \textbf{173},
  677 (1921).

\bibitem{con} B. D. Chowdhury, Pramana \textbf{70}, 593 (2008).
  E. T. Akhmedov, V. Akhmedova and D. Singleton, Phys. Lett. B
  \textbf{642}, 124 (2006).  E. T. Akhmedov, V. Akhmedova, T. Pilling
  and D. Singleton, Int. J. Mod. Phys. A \textbf{22}, 1705 (2007).
  V. Akhmedova, T. Pilling and A. de Gill and D. Singleton,
  arXiv:0804.2289, (2008).  T. K. Nakamura, arXiv:0706.2916 (2007).
  P. Mitra, Phys. Lett. B \textbf{648}, 240 (2007).

\bibitem{others} K. Srinivasan and T. Padmanabhan, Phys. Rev. D
  \textbf{60}, 24007 (1999).  S. Hemming and E. Keski-Vakkuri,
  Phys. Rev. D \textbf{64}, 044006 (2001).  S. Shankaranarayanan,
  T. Padmanabhan and K. Srinivasan, Class. Quant. Grav. \textbf{19},
  2671 (2002).  A. J. M. Medved, Phys. Rev. D \textbf{66}, 124009
  (2002).  E. C. Vagenes, Phys. Lett. B \textbf{559}, 65 (2003).
  T. Padmanabhan, Mod. Phys. Lett. A \textbf{19}, 2637 (2004).
  A. J. M. Medved and E. C. Vagenas, Mod. Phys. Lett. A \textbf{20},
  2449 (2005).  M. Arzano, A. J. M. Medved and E. C. Vargenas, JHEP
  \textbf{0509}, 037 (2005).  Q. Q. Jiang, H. L. Li, S. Z. Yang and
  D. Y. Chen, Mod. Phys. Lett. A \textbf{22}, 891 (2007).

\bibitem{niel2} S. A. Haywood, Phys. Rev. Lett. \textbf{96}, 031103
  (2006).  A. B. Nielsen, arXiv:0711.0313 (2007). A. B. Nielsen,
  arXiv:0804.4435 (2008).

\bibitem{visser} M. Visser, Int. J. Mod. Phys. D \textbf{12}, 649 (2003).

\bibitem{di} R. Di Criscienzo, M. Nadalini, L. Vanzo, S. Zerbini and
  G. Zoccatelli, arXiv:0707.4425 (2007).

\bibitem{niel1} A. B. Nielsen, arXiv:0802.3442 (2008).

\bibitem{stretch} L. Susskind, L. Thorlacius and J. Uglum,
  Phys. Rev. D \textbf{48}, 3743 (1993).

\bibitem{fro} V. Frolov, P. Sutton and A. Zelnikov, Phys. Rev. D
  \textbf{61}, 024021 (1999).

\bibitem{dav3} P. C. W. Davies, S. A. Fulling and W. G. Unruh,
  Phys. Rev. D \textbf{13}, 2720 (1976).

\bibitem{singh} T. P. Singh and C. Vaz, Phys. Lett. B \textbf{481}, 74 (2000).


\end{thebibliography}
\end{document}